\documentclass[12pt]{iopart}

\usepackage{graphicx}

\begin{document}

\input epsf.sty

\title{Extinction in four species cyclic competition}

\author{Ben Intoy and Michel Pleimling} \address{Department of Physics, Virginia Tech, Blacksburg, Virginia 24061-0435, USA}

\begin{abstract}
When four species compete stochastically in a cyclic way, the formation of two teams of mutually neutral partners is observed. In this
paper we study through numerical simulations the extinction processes that can take place in this system both in the well mixed
case as well as on different types of lattices. The
different routes to extinction are revealed by the
probability distribution of the domination time, i.e. the time needed for one team to fully occupy the system.
If swapping is allowed between neutral partners, then the probability distribution is dominated by very long-lived
states where a few very large domains persist, each domain being occupied by a mix of individuals from species
that form one of the teams. Many aspects of the possible extinction scenarios are lost when only considering
averaged quantities as for example the mean domination time. 
\end{abstract}

\maketitle

\section{Introduction}

In recent years biodiversity and species extinction in ecological networks \cite{May74,Smith74,Sole06} have yielded an increased interest
among statistical physicists \cite{Sza07,Fre09}, due to the many novel, and often unexpected, features that emerge when
going beyond the mean-field treatment of the simplest predator-prey models. Already simple modifications, like adding stochastic
effects \cite{McK95} and/or a spatial environment \cite{Mob06,Mob07,Tau12}
to the standard Lotka-Volterra model, can change markedly species coexistence and extinction. Similarly, going beyond a simple predator-prey
relationship by allowing for more than two species yields interesting new scenarios that have been the focus of a range
of recent studies \cite{Fra96a,Fra96b,Fra98,Kob97,Pro99,Sza01,Sza01b,Tse01,Sat02,
Ker02,Sza04,Kir04,Sza05,He05,Rei06,Sza07b,Sza07c,Per07,Rei07,Rei07a,Sza08,Sza08a,Cla08,Pel08,Rei08a,Ber09,Ven10,Cas10,Shi10,
And10,Wan10,Mob10,He10,Win10,Nob11,Dur11,He11,Rul11,Wan11,Nah11,Jia11,Pla11,Dem11,Zia11,He12,Van12,Don12,Dob12,
Juu12,Lam12,Jia12,Ada12,Juu12a,Dur12,Rom12,
Lut12,Ave12a,Ave12b,Rom13,Kne13,Rul13,Szs13,Ave13}.

In most situations biodiversity is only a transient phenomenon in systems characterized by species competition. 
Stochastic effects tend to favor species extinction in finite
populations, and this even for cases where the mean-field rate equations predict ever-lasting coexistence in the form
of population oscillations \cite{Rei08a,Ber09,Dur12,Par09}. The end of biodiversity can be captured by the extinction time
defined as the time until the first species dies out. A closely related notion is that of fixation time in evolutionary
game theory which measures the time for one strategy to take over the entire population \cite{Ant06}.

Extinction times have been studied extensively in situations where three species compete in a cyclic way. 
Of special interest are cases with symmetric rates, and we will focus on this in the following.
When the total number of individuals is constant, the deterministic dynamics is characterized by neutrally
stable, closed orbits (due to the existence of a conserved quantity \cite{Ant06,Rei07,Cre09}).
For such a case of neutrally stable coexistence one expects that in the stochastic model
the mean extinction time scales algebraically with the system size $N$. In \cite{Rei06,Cla08,Dob12}
it was shown that for the well mixed stochastic rock-paper-scissors model the mean extinction time indeed increases linearly with $N$.
For the related May-Leonard model \cite{May75}, however, where the total number of individuals is a stochastic variable, 
the mean extinction time in the well mixed case increases logarithmically with $N$ \cite{Rei07a}. This is readily
understood by noting that in that case the deterministic dynamics yields heteroclinic orbits that are unstable against
demographic fluctuations \cite{Fre09,Mob10}. The situation gets more complicated when considering spatially extended
systems. In one space dimension it is found that for the rock-paper-scissors case the extinction time increases algebraically with the
system size, but with an exponent that is larger than one. The distribution of extinction times is found to have a broad tail so
that the average is dominated by rare events of long lasting coexistence \cite{He11}. In two space dimensions
the well mixed results hold for both models for large particle mobility, as this provides an effective mixing mechanism.
Below a certain mobility threshold, however, the mean extinction time increases exponentially with the system size \cite{Rei07a}.
The consideration of spatially inhomogeneous reaction rates does not markedly change these scenarios \cite{He10,He11}.
Extinction times have also been discussed recently for a three species cyclic model where the rock-paper-scissors
and May-Leonard schemes are combined \cite{Rul13}.

It is important to note that three species in cyclic competition form a very special case, as here every species interacts
with every other species through a predator-prey relationship. When considering more than three species one has the
possibility to have mutually neutral, i.e. non-interacting, species. Obviously in a real ecological environment mutually
neutral species are common.

Whereas recently an increasing number of studies 
focused on cases with four or more species \cite{Fra96a,Fra96b,Kob97,Fra98,Sza01,Sza01b,Sat02,Sza04,Sza05,He05,Sza07b,Per07,Sza08,Cas10,
Nob11,Dur11,Zia11,Van12,Dob12,Dur12,Rom12,Lut12,Ave12a,Ave12b,Rom13,Kne13,Ave13}, not much is known about the corresponding extinction times.
In \cite{Dob12} the mean extinction time for the cyclic four species case is shown to scale linearly with the system size in the
well mixed situation without spatial dependence. The size dependence of the mean extinction time is briefly discussed in \cite{Kne13}
for a well mixed five species model with a more complicated interaction scheme.

In this paper we use numerical simulations to study in a systematic way extinction events in the cyclic 
four species case. In contrast to most of the aforementioned
studies we do not solely discuss average quantities. Instead our focus is on probability distributions which are
found to be non-trivial and to reflect the different stages of the competition between the species. 
We study the well mixed situation as well as the regular one- and two-dimensional lattices.
We also present results for the four species model on the Sierpinski triangle
which has a fractal dimension of approximately 1.585. Starting from a completely disordered
initial state, we find that for spatial systems the probability distribution exhibits two different long time regimes. These two
regimes correspond to two very different extinction scenarios.

The paper is organized in the following way. In the next section we introduce the different versions of the four species model 
discussed in this paper. In section
3 we present our results for the different cases, namely the  well mixed system as well as various 
spatially extended systems: the line, the Sierpinski triangle, and the square lattice. The focus of our study is on the
probability distribution, which for the spatially extended systems is characterized by two different time scales. 
Section 4 gives our conclusions.

\section{Model and geometries}

We consider four species that undergo predator-prey interactions in a cyclic way. Calling the different species $A$, $B$,
$C$, and $D$, these interactions can be
cast symbolically in the form of reactions:
\begin{eqnarray} 
A+B && \stackrel{\mu_a}{\to} A+A  \nonumber \\
B+C && \stackrel{\mu_b}{\to} B+B  \nonumber \\
C+D && \stackrel{\mu_c}{\to} C+C  \nonumber \\
D+A && \stackrel{\mu_d}{\to} D+D \nonumber
\end{eqnarray}
In this four species model we have
the presence of two pairs of neutral species, namely ($A, C$) and ($B, D$), that do not have a predation-prey relationship. This
is an important difference with the three species case where every species interacts with every other. 

This reaction scheme yields the following mean-field or rate equations for the species concentrations (see \cite{Cas10,Dur11}
for a detailed discussion of the mean-field results):
\begin{eqnarray}
\partial_t a & = & [\mu_a b-\mu_d d]a ~~~,~~~\partial_t b ~ = ~ [\mu_b c-\mu_a a]b \nonumber \\
\partial_t c & = & [\mu_c d-\mu_b b]c ~~~,~~~\partial_t d ~ = ~ [\mu_d a-\mu_c c]d \nonumber
\end{eqnarray}
which can be cast in the form
\begin{eqnarray}
\partial_t [\mu_b\textrm{ln}a+\mu_a\textrm{ln}c] ~ = ~ \lambda d~~~,~~~
\partial_t [\mu_c\textrm{ln}a+\mu_d\textrm{ln}c] ~ = ~\lambda b \nonumber \\
\partial_t [\mu_c\textrm{ln}b+\mu_b\textrm{ln}d] ~ = ~ -\lambda a~~~,~~~
\partial_t [\mu_d\textrm{ln}b+\mu_a\textrm{ln}d] ~ = ~-\lambda c \nonumber
\end{eqnarray}
with $\lambda\equiv k_ak_c-k_bk_d$. Adding and subtracting these equations one discovers that 
the quantity
\begin{equation}
Q\equiv \frac{a^{\mu_b+\mu_c}c^{\mu_d+\mu_a}}{b^{\mu_c+\mu_d}d^{\mu_a+\mu_b}}
\end{equation}
shows a very simple dependence on time:
\begin{equation}
Q(t)=Q(0)e^{\lambda t}~.
\end{equation}
Notice that if $\lambda=0$ then $Q(t)$ is a constant of motion for the deterministic evolution,
determined by the initial concentrations. In that case
the concentrations display periodic, ever-lasting oscillations.

In this work we focus on the interesting case where all predation rates are equal, i.e. $\mu_a = \mu_b = \mu_c = \mu_d = \mu$,
which yields $\lambda = 0$.
For this case the mean-field predictions of species coexistence and population oscillations markedly differ from 
the domination by one of the partner pairs encountered in the stochastic evolution.
See \cite{Dur12} for an in-depth comparison
between mean-field approximation and stochastic evolution.

We consider in the following both the well mixed situation, where every agent interacts with every other, as well as lattice systems with
single site occupation, where individuals located on neighboring lattice sites can prey on each other. We do not allow for
empty sites, which entails that the total number of individuals in the system
is conserved. We consider mobile individuals that can swap places with one of their neighbors.
In case these two individuals have a predator-prey relationship, then the swapping takes place with rate $\sigma$ thus that $\sigma + \mu =1$.
It can also happen that the two agents on neighboring sites are from two mutually neutral species. If that is the case, then swapping is allowed
to take place with rate $\sigma_n$ (the index $n$ indicates that this is the rate for the swapping of neutral partners). 
We discuss below the two cases where $\sigma_n = \sigma$ or $\sigma_n = 0$.

\begin{figure}[h]
\centering
\includegraphics[width=0.75\textwidth]{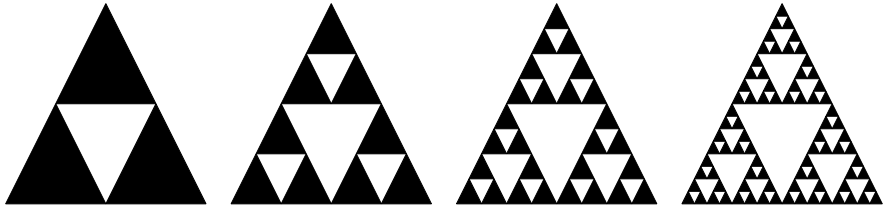}
\caption{Sierpinski triangle after $m$ iterations. From left to right: $m = 1, 2, 3, 4$}
\label{fig1}
\end{figure}

Besides studying regular one- and two-dimensional lattices, both with open and periodic boundary conditions, we also simulate our four species
game on a fractal. The fractal we discuss in the following is the Sierpinski triangle with Hausdorff dimension $\log(3)/\log(2) \approx
1.585$. This fractal is obtained by repeatedly dividing each triangular plaquette into four solid triangular plaquettes and removing the center.
Since we are interested in finite lattices, we define the depth $m$ of a Sierpinski triangle to be the number of 
dividing and removing iterations. Figure \ref{fig1} shows the resulting lattices after the first few iterations. Every lattice point has
three neighbors, with exception of the three outermost vertices which have only two. For the Sierpinski triangle of depth $m$
the total number of lattice sites is given by $3^m$.
This lattice structure implies a hierarchy of bottlenecks, i.e. lattice points through which one has to go when crossing
from one part of the system to another. From this point of view the Sierpinski triangle is intermediate between the chain, where every lattice
site can be viewed to be such a bottleneck, and the regular square lattice where one has many equivalent paths between different parts of the system.

If not stated otherwise we prepare the system in an initial state where every lattice site is occupied with the same probability by any one of the
four species. Reactions and exchanges are then taking place
between particles located on neighboring lattice sites. For every update we select a pair of neighboring sites and apply the rules given above.
We increase time by one unit after $N$ proposed updates where $N$ is the number of sites/individuals in the system.
We stop the simulation when one neutral species pair, either $(A,C)$ or $(B,D)$, is occupying the whole system. The time at which this happens
is then taken as our domination time. This domination time is closely related to the extinction time
at which the first species goes extinct. Indeed, once a species dies out, the prey of its prey, i.e. its partner, will also be
dismissed very rapidly.

In order to understand our results for the domination time, we also study the average domain size of individual species as well
as the average domain size of neutral species pairs. As shown in \cite{Rom12}, the four species on a lattice tend to arrange themselves 
into domains, yielding ultimately a coarsening process of domains occupied by the mutually neutral species pairs. Studying these
two different domain sizes allows us to 
relate the different phases of the coarsening process to the different regimes displayed by the probability distribution of the
domination time.

\section{Domination times}

\subsection{The well mixed case}

Let us start our discussion of the domination time $\tau$ with the simple case of a well mixed system without
an underlying lattice. It has been shown in \cite{Dob12} that in the well mixed four species case the mean extinction
time increases linearly with the system size. In our simulations two particles
are picked at random and allowed to perform a reaction following the scheme discussed previously. Since there
is no spatial structure, we set the predation rate $\mu = 1$ (a different value of $\mu$ only rescales time). Initially, every one of the $N$ sites
in the system is assigned to one of the four species with equal probability. 

\begin{figure}[h]
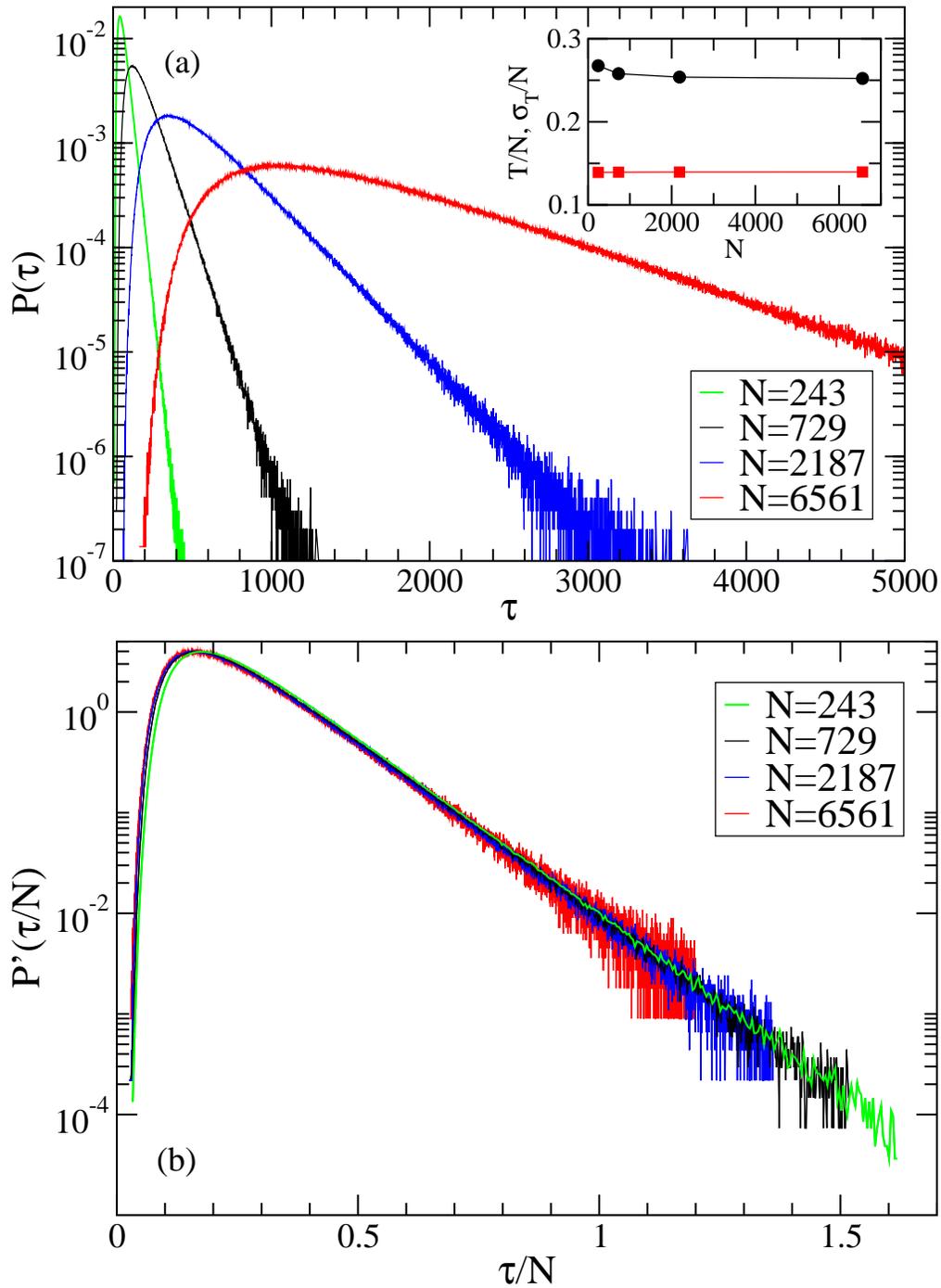

\centerline{\epsfxsize=5.25in\ \epsfbox{figure2a.eps}}
\vspace*{0.2cm}
\centerline{\epsfxsize=5.25in\ \epsfbox{figure2b.eps}}
\caption{(a) Probability distribution of the domination time for different system sizes $N$ in the well mixed case
with $\mu =1$. The inset shows that both the mean domination time $T$ (black circles) and its standard deviation $\sigma_T$ (red squares) increase
linearly with the system size for not too small systems. (b) Scaling of the probability distribution with system size $N$, see
equation (\ref{eq:Pscal}). 
The probability distributions are obtained from at least seven million
independent simulations.
}
\label{fig2}
\end{figure}

The probability distribution $P(\tau)$ of the domination time is shown in figure \ref{fig2}a for different system
sizes. Starting from a random mix, a non-zero minimum amount of time is needed for one of the neutral species pairs to
dominate. Consequently, the probability distribution rises for small times until it reaches a maximum. The decaying part
is well described by a shifted exponential distribution
\begin{equation} \label{eq:P}
\tilde{P}(\tau) = \beta \, e^{-\beta \, \left( \tau - \tau_0 \right)} \Theta(\tau - \tau_0)
\end{equation}
where $\Theta$ is the step function, with the shift $\tau_0$ and the parameter $\beta$.
This exponential distribution is a consequence of the fact that the system essentially performs an unbiased random walk in 
configuration space. The extinction of a species (which rapidly yields the extinction
of its partner) can then be viewed as a Poisson process described by an exponential
distribution. Equation (\ref{eq:P}) captures this behavior while taking into account the fact
that it takes some time for a species pair to dominate when starting from a random
mix.

From the probability distribution we obtain the mean domination time $T = \langle \tau \rangle = \sum\limits_{\tau =0}^\infty
\tau P(\tau)  \approx 0.25 ~N$ and the standard deviation $\sigma_T = \sqrt{\langle \tau^2 \rangle - \langle \tau \rangle^2} \approx 0.14~N$, see
the inset of figure \ref{fig2}a. From equation (\ref{eq:P}) also follows the relationships
$\tau_0 = T - \sigma_T$ and $\beta = 1/\sigma_T$, from which one immediately deduces that the shift $\tau_0$ is a linear function of $N$,
whereas the parameter $\beta$ varies inversely proportional to $N$.

It also follows from the shifted exponential distribution (\ref{eq:P}) that data from different system sizes should collapse on a 
master curve when scaled properly. Indeed, exploiting directly the size dependence of the mean, $T = T_\infty N$, and the standard deviation,
$\sigma_T = \sigma_\infty N$, as well as their relationships with $\tau_0$ and $\beta$, equation (\ref{eq:P}) can be recast in
the scaling form
\begin{equation} \label{eq:Pscal}
\tilde{P}'(\tau/N) = \frac{1}{\sigma_\infty} e^{\tau_\infty/\sigma_\infty} \, e^{- \frac{1}{\sigma_\infty} \left( \tau/N \right) } \, \Theta\left[\left( \tau/N \right) 
- \tau_\infty \right]
\end{equation}
with $\tau_\infty = T_\infty - \sigma_\infty$. As verified in figure \ref{fig2}b this scaling indeed works very well 
for the well mixed case.

\subsection{Lattice systems}

\begin{figure}[h]
\centerline{\epsfxsize=4.25in\ \epsfbox{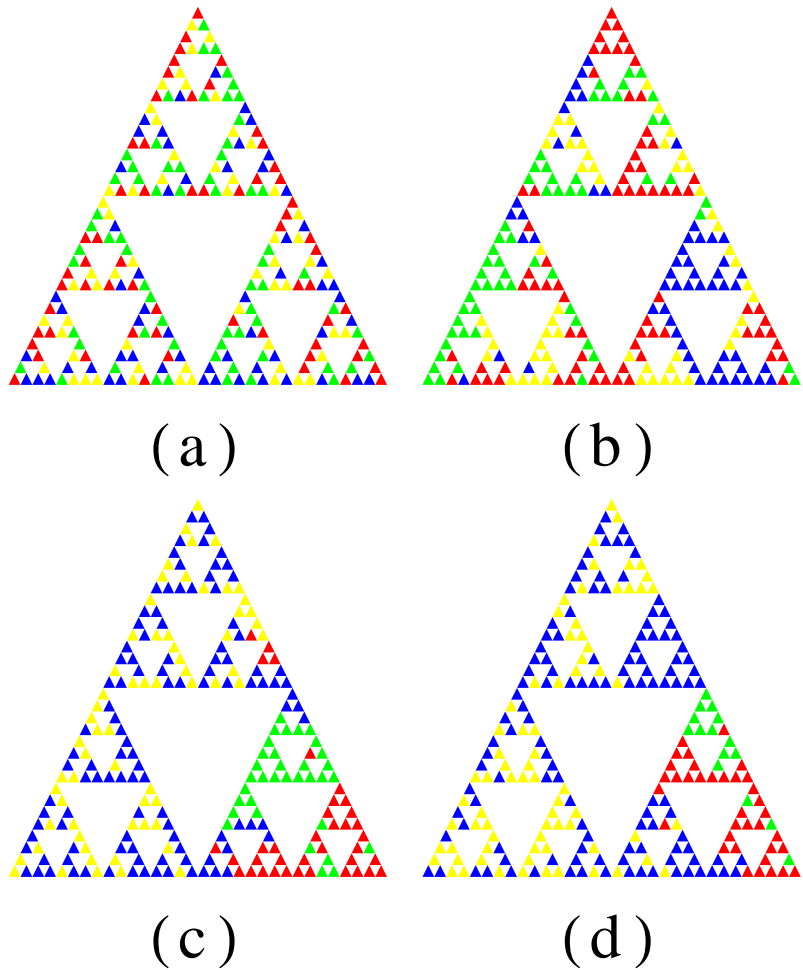}}
\caption{Snapshots of a simulation on the Sierpinski triangle with $\mu = 0.8$ and $\sigma = \sigma_n = 0.2$. The system
size is 243 (iteration depth $m=5$). The snapshots are taken at times (a) $t=0$ (initial state), (b) $t=9$, (c) $t=510$, 
and (d) $t=5430$ (just before the yellow-blue
team dominates the system). Time is measured in Monte Carlo steps.
}
\label{fig3}
\end{figure}

The four species cyclic game on the lattice is characterized by the competition between the two different teams
composed of mutually neutral partners. As a result, coarsening of domains occupied by the different teams sets in \cite{Rom12}.
These coarsening domains are compact and rarely contain individuals from the enemy team. In two space dimensions the domain
boundaries are not very sharp, due to relentless reactions (predation and swapping) taking place at these boundaries. See 
\cite{Rom12} for some configurations in one and two space dimensions. In figure \ref{fig3} we show some snapshots from a typical
run on the Sierpinski triangle. After preparing the system in a disordered initial state (a), small domains are rapidly formed (b),
followed by a phase of domain growth (c). Due to the presence of bottlenecks that separate the lattice in different parts, one often
observes that different parts of the lattice are occupied by different teams. This
blocked situation can last for quite some time. For example, for the run shown in figure \ref{fig3} not much is happening between the 
snapshots (c), taken at time $t=510$, and (d), taken at $t=5430$. Between these two times there are many excursions into enemy territory
that fail. Eventually, one of these excursions is successful, and one of the teams takes over the whole lattice. The configuration
(d) shows the beginning of that final excursion, just before the red and green species go extinct. 

\begin{figure}[h]
\centerline{\epsfxsize=5.25in\ \epsfbox{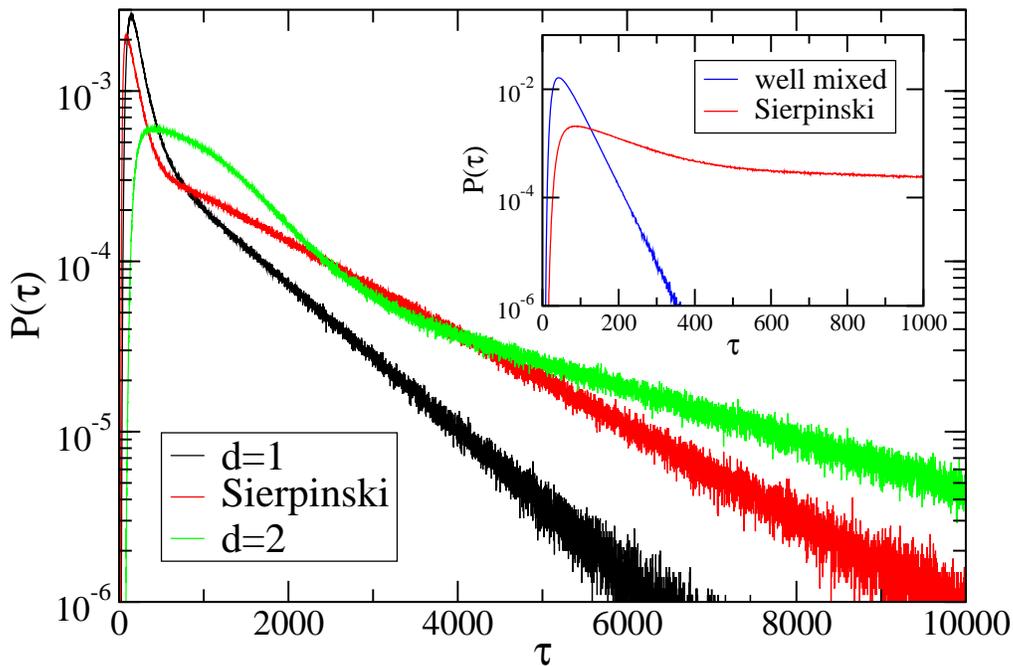}}
\caption{Typical probability distributions of the domination time for lattice systems. 
The one-dimensional chain contains $N=81$ sites, with $\mu = 0.1$, whereas the Sierpinski triangle is formed
by $N= 243$ sites, with $\mu = 0.5$. Finally, the square has $N = 45 \times 45$ sites, with $\mu = 0.6$. The swapping rates 
are always given by $\sigma = \sigma_n = 1 - \mu$. Inset: comparison of the well mixed (see figure \ref{fig2}) and Sierpinski cases with the same
number of individuals $N = 243$.
}
\label{fig4}
\end{figure}

In order to investigate species extinction in these lattice systems, we restrict ourselves to small systems. For the line and the Sierpinski
triangle the sizes range from $N=16$ to $N=2197$, whereas in two dimensions we consider lattices from $N = 6 \times 6$ to $N = 108 \times 108$
sites. For the studied systems we typically did millions of independent runs. In one and two space dimensions we focus
on closed systems. We have studied
in a similar way one- and two-dimensional systems with periodic boundary conditions. However, as the results 
for periodic boundary conditions are very similar to those
obtained in closed systems (there are of course quantitative differences, but qualitatively the studied quantities behave in similar ways
for the different boundary conditions), we refrain from discussing in detail our results obtained for periodic boundaries.

Figure \ref{fig4} shows for the three different lattice types typical probability distributions of the domination time. 
Whereas in all cases we observe the initial raise for small times and an exponential decay for long times, an additional intermediate regime is
observed for the lattice systems that is absent in the well mixed case, compare with figure \ref{fig2}. Indeed, after the maximum 
a first exponential decaying regime is encountered before a crossover to the final exponential decay sets in. This final decay
is much slower than for the well mixed case, see the inset. 
This intriguing shape of the probability distribution
indicates the presence of two different time scales. As discussed in the following, these two different time scales correspond to two 
different routes to extinction.

\begin{figure}[h]
\centerline{\epsfxsize=3.50in\ \epsfbox{figure5a.eps}}
\vspace*{0.2cm}
\centerline{\epsfxsize=3.40in\ \epsfbox{figure5b.eps}}
\vspace*{0.2cm}
\centerline{\epsfxsize=3.50in\ \epsfbox{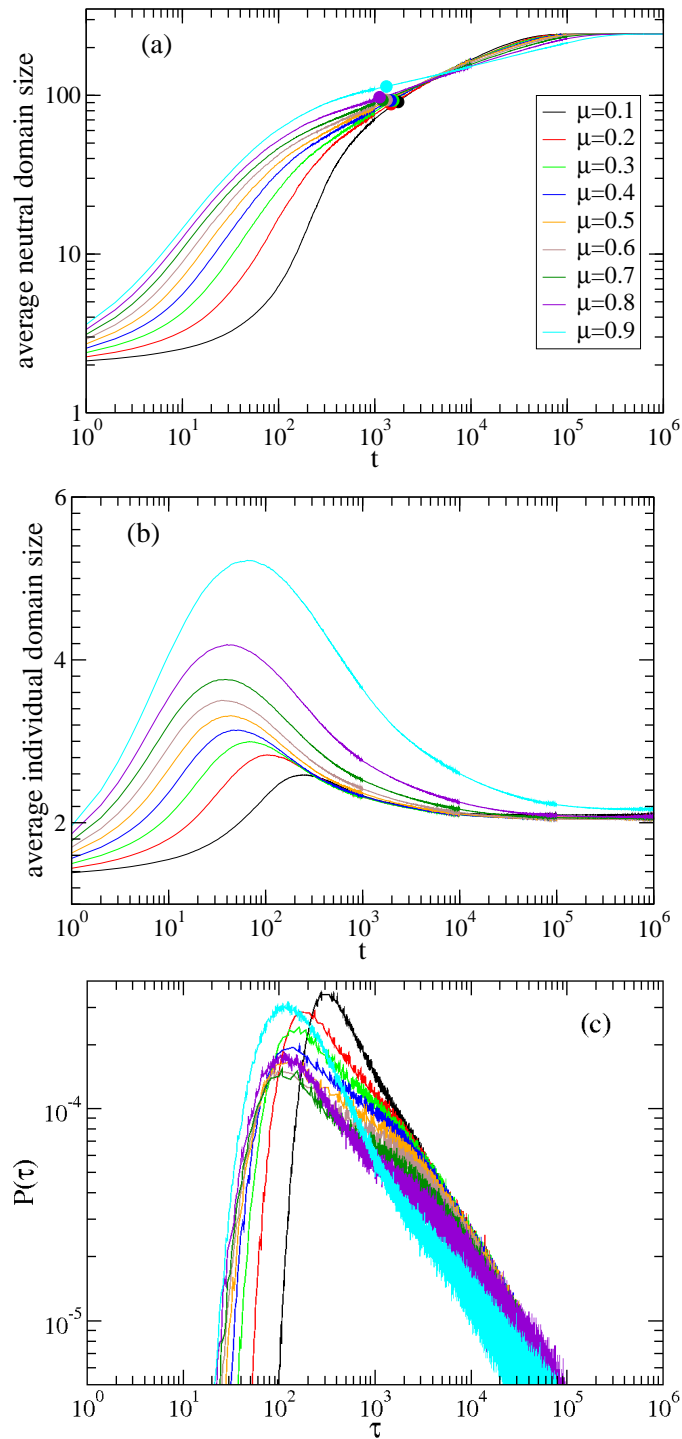}}
\caption{(a) Average neutral domain size, (b) average individual domain size, and (c) domination time distribution
 for the line composed of $N =243$ sites.
Date for different predation rates $\mu$ are shown, where the swapping rates are given by $\sigma = \sigma_n = 1- \mu$. The
dots in (a) indicate the domination times where one passes from one exponentially decaying regime to the next, see the change
of slope in (c).
The data result from averaging over 10000 independent runs.
}
\label{fig5}
\end{figure}

Possible hints at the origin of these two time scales can be found in the average domain size. Due to the presence of mutually neutral
species we need to distinguish between
the average size of domains that contain individuals of only one species and the average size of the larger domains 
formed by neutral partners. The typical time dependence of these sizes are shown in figure \ref{fig5} for the line containing 243
sites. Analysis of figures \ref{fig5}a and \ref{fig5}b reveals three different growth regimes, in close agreement with the 
features in the probability distributions shown in figure \ref{fig5}c.

As the initial preparation is in a random state, many small single species domains composed of only very few individuals
are formed initially. 
Some of them will keep growing at the expense of others. In this stochastic process species are eliminated locally, 
which can lead in some cases to the global extinction of one of the neutral pairs, hence the increase of the domination time
probability visible in figures \ref{fig4} and \ref{fig5}c. The local dismiss of species will be followed by an increase of the encounters
of neutral species domains, resulting in the strong increase of the neutral domain size seen in figure \ref{fig5}a as well as in 
a decrease of the domination probability function. Concomitantly the neutral species start to diffuse into each other, yielding the
decrease of the individual domain size seen in figure \ref{fig5}b. Once neutral domains are well mixed, it becomes very 
difficult for a team to take over part of the system occupied by the competing team, as every species is confronted with a 
mixture of predators and preys. This yields a much slower domain growth, as witnessed by the change of slope in figure \ref{fig5}a,
and the emergence of long-lasting transients which are revealed by the change of slope in figure \ref{fig5}c. The simultaneity
of these two events becomes obvious when including in figure \ref{fig5}a, see the circles, the domination time where the transition
between the two exponential regimes in the probability distribution takes place. 

Let us add that the presence of the different regimes is a generic property of finite lattice systems, independent of the 
system size $N$ and the values of the reaction
rate $\mu$, provided that a mechanism is in place that allows to well mix partner pair domains. 
Only in the limit $\mu \ll 1$, when particles mainly swap places, is coarsening absent, and no long lived
states are encountered in the system.

\begin{figure}[!h]
\centerline{\epsfxsize=5.25in\ \epsfbox{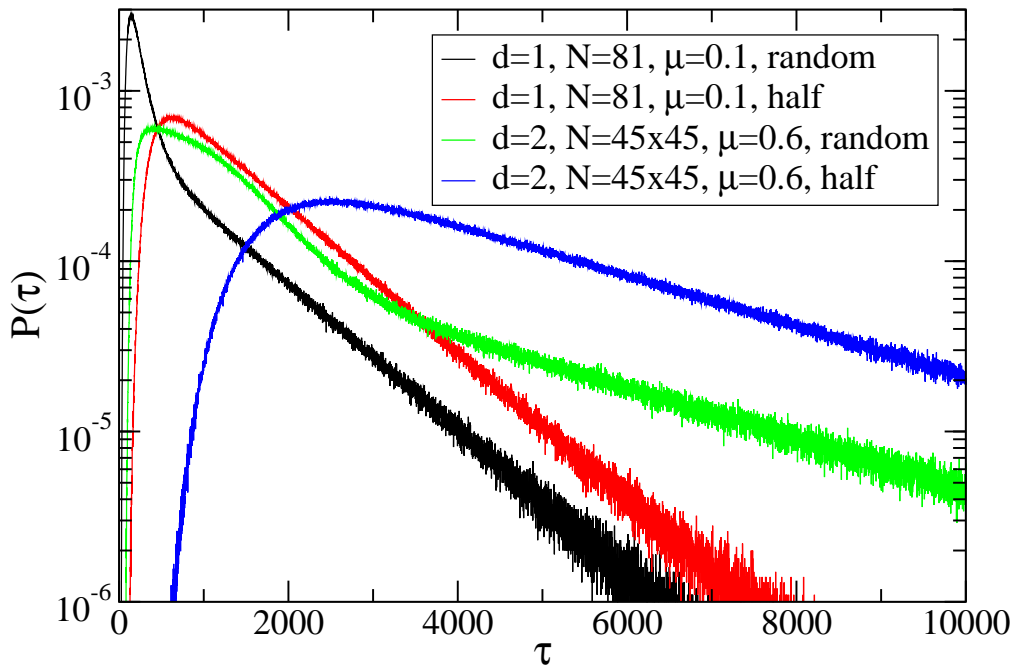}}
\caption{Probability distributions of the domination time for the line and the square lattice with different initial conditions: random initial
condition as well as an initial state where the different teams each occupy half of the system and are arranged in such a way that
every individual is surrounded by members of the partner species. The decay rate of the exponential decay is independent of the initial state 
which indicates that these long-lived states are exclusively due to the competition of a few very large well mixed domains.
}
\label{fig6}
\end{figure}

\begin{figure}[h]
\centerline{\epsfxsize=5.25in\ \epsfbox{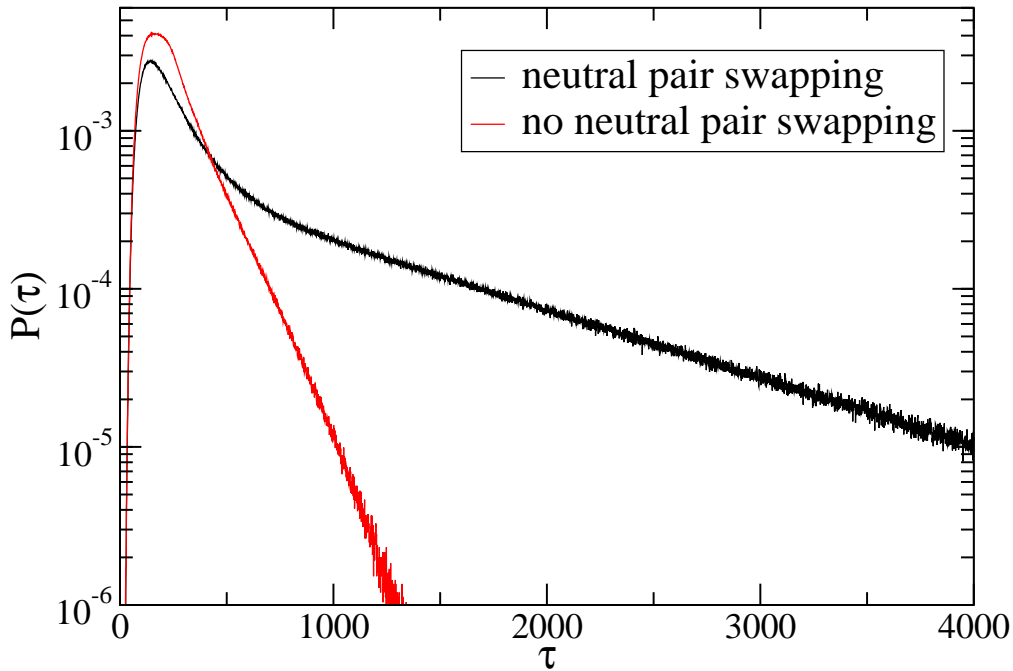}}
\caption{Probability distributions of the domination time for the line with and without neutral pair swappings. 
The long-lived states are absent without an efficient mixing mechanism between neutral partners. The system
size is $N=81$ and $\mu = 0.1$. The probability distributions result from more than 12 million independent runs.
}
\label{fig7}
\end{figure}

Figures \ref{fig6} and \ref{fig7} provide further support for our interpretation of the involved processes. We first check in
figure \ref{fig6} whether these long lived states are indeed due to the presence of well mixed neutral domains. For that
we prepare the system in such a way that every half of the system is occupied by one of the partner pairs. In every half
the two allying species are mixed completely by having every individual surrounded by members of the partner species (with
the exception of the particles located at the interface between the two halves). For the line,
this yields an initial state of the form: $\cdots ACACACBDBDBD \cdots$, whereas for the square lattice we have a checkerboard of $A,C$ and
$B,D$ particles in the different halves. Figure \ref{fig6} compares the probability distributions for the domination time obtained for these 
initial states with those that follow from random initial conditions. For both initial conditions the decay rate of the long-lived
states is the same, indicating that these states are indeed due to the competition of few very large well mixed domains.

In order to have well mixed domains composed of mutually neutral species an efficient mechanism for the mixing of these
species needs to be in place. This is realized in our system through the swapping of neutral species with the rate $\sigma_n =
1-\mu$. Without this swapping mechanism we expect that the long-lived states should be completely absent, whereas the earlier stages
of the coarsening process, that do not rely on this swapping mechanism, should be very similar. This is indeed the case, see
figure \ref{fig7} where we compare for the line with $N=81$ sites and $\mu = 0.1$ the probability distributions with and
without neutral pair swappings.

The probability distributions of the domination time reveal many interesting features of the emerging spatio-temporal
correlations due to the cyclic competition between species. It is obvious that much information is lost when only
looking at the mean domination time (extinction time) as has been done in most of the recent studies. Still, because
of the focus in the literature on this mean time, it is of interest to also discuss that quantity for
our lattice systems.

\begin{figure}[h]
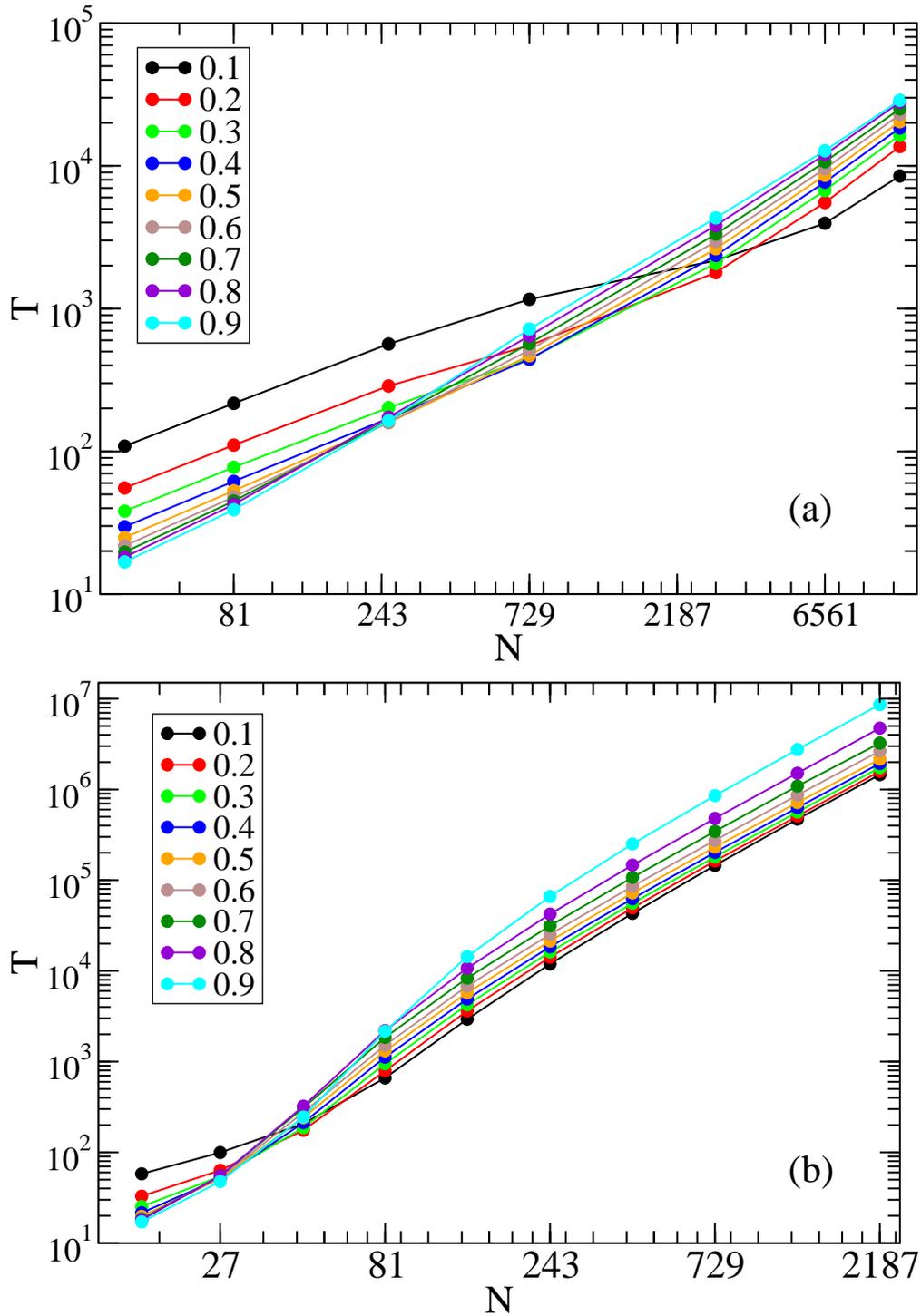

\centerline{\epsfxsize=5.25in\ \epsfbox{figure8a.eps}}
\vspace*{0.2cm}
\centerline{\epsfxsize=5.25in\ \epsfbox{figure8b.eps}}
\caption{The mean domination time $T$ versus the total number of sites $N$ for (a) the square with closed
boundaries, with system sizes ranging from $N= 36 = 6 \times 6$ to $N = 11664 = 108 \times 108$, and (b)
the line. The different data sets correspond to the different values of $\mu$ given in the legend. The swapping
rates are $\sigma = \sigma_n = 1- \mu$. The data result from averaging over at least 100000 independent runs.
}
\label{fig8}
\end{figure}

Figure \ref{fig8} shows the size dependence of the mean domination time $T$ for 
different predation rates $\mu$ and swapping rates $\sigma=\sigma_n = 1-\mu$. The data for the square, see
figure \ref{fig8}a, clearly shows a crossover between two regimes. For small system sizes the system behaves like
a well mixed system, and the mean domination time increases approximately linearly with the system size. However, for
the largest system sizes studied in this work spatial effects become important, and $T$ varies algebraically: $T \sim N^\alpha$,
with an exponent $\alpha \approx 1.45$. The same crossover is also present in the one-dimensional system, see
figure \ref{fig8}b, but with a different exponent for the larger systems. Fitting the data for the largest three 
system sizes to a power law yields an effective
exponent $\alpha \approx 2.10$.

\begin{figure}[h]
\centerline{\epsfxsize=5.25in\ \epsfbox{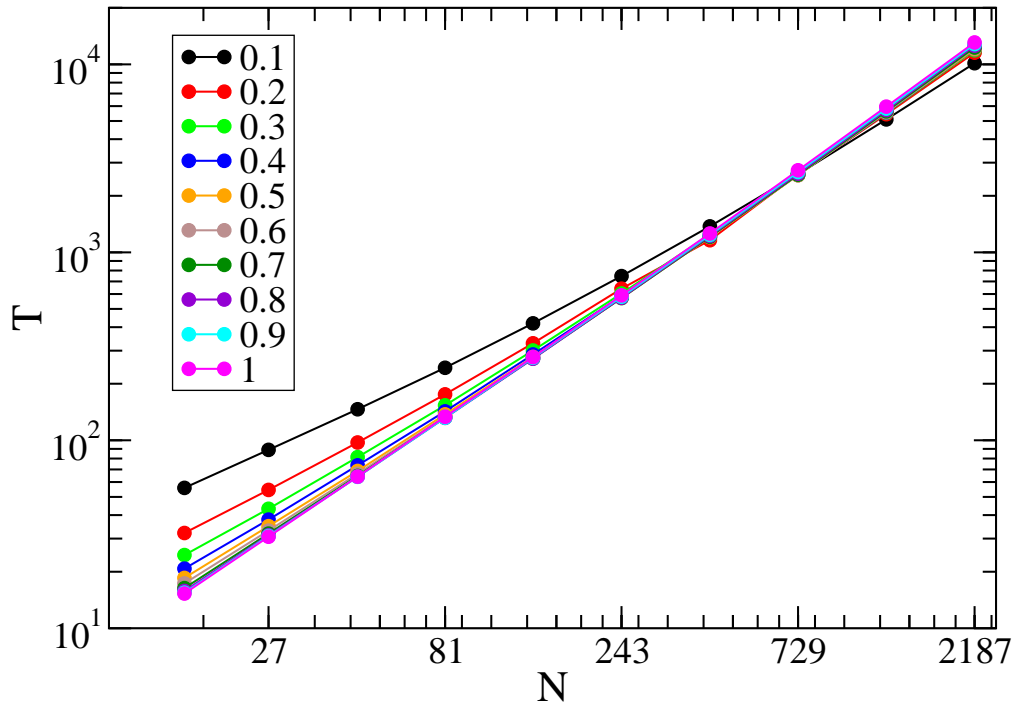}}
\caption{The same as in figure \ref{fig8}b, but now without swapping between neutral partners, i.e. $\sigma_n = 0$.
The data result from averaging over at least 100000 independent runs.
}
\label{fig9}
\end{figure}

Finally, we note that the same crossover is also observed in absence of neutral partner swappings, see 
figure \ref{fig9}. However, this crossover is much more gradual, presumably due to the absence of the long-lived
states where well mixed domains coarsen very slowly. 
Another difference between this case and that of figure \ref{fig8}b is that in absence
of neutral swappings the mean domination times increases much slower for larger system sizes. Indeed, for
large $N$ we obtain $T \sim N^\alpha$ with $\alpha \approx 1.40$.

\section{Discussion and conclusion}
Stochastic evolution of a finite population of predators and preys eventually ends in species extinction and loss of
biodiversity. In many instances very different routes to species extinction are possible, where the different extinction
scenarios might prevail at different stages of the time evolution. 

In this paper we have studied the extinction processes that are encountered in a system of four species that compete
in a cyclic way. In contrast to the much studied three species models where every species is in a predation-prey
relationship with every other, the four species cyclic competition is a simple case where some species are mutually
neutral. As a result the four species tend to form two alliances that compete against each other.
We study this model both in the well mixed case without any spatial dependence as well as on a variety of lattices:
the regular one- and two-dimensional lattices as well as the fractal Sierpinski triangle.
In the lattice systems our individuals are mobile and they can swap places with their neighbors. These swappings can take
place between predators and preys but also between mutually neutral partners.

Our study focuses on the probability distribution of the domination time, which is the time at which one of the teams
completely fills the lattice. This probability distribution is a complicated function, with various regimes
which can be related to different extinction processes. In presence of neutral pair swappings the probability distribution
exhibits a crossover between two different exponential decays. The earlier regime corresponds to extinctions taking place
during the coarsening of domains that contain mostly one species. The second regime, characterized by very broad tails,
results from extremely long-lived states that are due to the competition of few large domains where the members of one of
the teams are well mixed. This yields a stalemate as every domain is surrounded by domains that contain
a mixture of preys and predators. We have verified this scenario through simulations where we
prepared the system in an initial state where each half is occupied by one team, with the team members occupying the lattice
in an alternate way. In absence of the swappings of neutral partners, which provide the efficient mixing mechanism inside the
domains occupied by a single team, these long-lived states are absent and the probability distribution does not exhibit
this marked crossover between different regimes.

It is worth pointing out that the most prominent features of the probability distribution are independent of the lattice type.
Thus in presence of neutral partner swappings
one observes for every lattice the emergence of long-lasting states characterized by well mixed large domains responsible
for the crossover between the two different regimes with exponential decay.

As the probability distributions are rather complicated, important information can be lost when looking at averaged quantities.
Thus the mean domination time $T$ as a function of system size $N$ behaves in a qualitatively similar way
with or without neutral pair swappings, even so the corresponding probability
distributions are markedly different. For a fixed value of the predation and swapping rates a crossover is observed from a well mixed
situation in small systems, where $T$ increases linearly with $N$, to a lattice dominated behavior for larger systems characterized by 
an algebraic increase $T \sim N^\alpha$ with an exponent $\alpha > 1$. The value of the exponent is found to depend on the dimension
of the lattice and on the presence or absence of neutral partner swappings.

Whereas the details of our results are specific to the four species cyclic model studied in this paper, our study also reveals
aspects that are important for other food webs characterized by competition between the different species. Indeed, in a
spatial environment these systems tend to yield alliances that result in more or less complicated coarsening processes
\cite{Ave12a,Ave12b,Rom13}. In some cases a complicated dynamics takes place inside the coarsening domains. In all these
cases one expects a rich variety of routes to extinction. It follows from our study that the detailed understanding
of these extinction processes warrants an in-depth study of the probability distributions. Relying exclusively on averaged
quantities will only allow to gain very superficial insights into species extinction in these systems.

\ack
This work was supported by the US National
Science Foundation through grant DMR-1205309. We thank Sven Dorosz for useful discussions.

\end{document}